\documentclass[10pt,conference]{IEEEtran}
\IEEEoverridecommandlockouts
\usepackage{cite}
\usepackage{amsmath,amssymb,amsfonts}
\usepackage{graphicx}
\usepackage{textcomp}
\usepackage{xcolor}
\usepackage{tabularx}
\usepackage{color, colortbl}
\usepackage[misc]{ifsym}
\usepackage{diagbox}
\definecolor{lime}{rgb}{0.88,2,10}
\renewcommand{\baselinestretch}{.95}
\usepackage{subcaption}
\usepackage{wrapfig}
\usepackage[linesnumbered,ruled,vlined]{algorithm2e}

\usepackage{xpatch}
\usepackage{url}
\usepackage{multirow}
\usepackage{multicol}
\usepackage{wrapfig}
\usepackage{mathrsfs}

\usepackage[export]{adjustbox}
\usepackage{nomencl}
\usepackage{siunitx}
\usepackage{blindtext}
\usepackage[T1]{fontenc}
\usepackage[spaces,hyphens]{xurl}
\usepackage{float}
\usepackage{booktabs}
\usepackage{verbatim}
\newcommand*{\Resize}[2]{\resizebox{#1}{!}{$#2$}}%

\SetAlFnt{\small}

\SetKwComment{Comment}{$\triangleright$\ }{}

\usepackage[utf8]{inputenc}
\usepackage{enumitem}

\SetKwInput{kwInit}{Input}
\SetKwInput{kwInita}{Output}
\def\BibTeX{{\rm B\kern-.05em{\sc i\kern-.025em b}\kern-.08em
    T\kern-.1667em\lower.7ex\hbox{E}\kern-.125emX}}

\usepackage[numbers,sort&compress]{natbib}
\usepackage{caption}
\newcommand{\red}[1]{\textcolor{red}{[#1]}}

\newcommand{\fref}[1]{Fig.~\ref{#1}}

\newcommand{\sref}[1]{Section~\ref{#1}}
\usepackage[misc]{ifsym}

\usepackage[outline]{contour}
\contourlength{0.2pt}
\contournumber{15}

\usepackage{pifont}

\usepackage[outline]{contour}
\contourlength{0.2pt}
\contournumber{15}

\usepackage{fancyhdr}
\fancypagestyle{mystyle}{
    \chead{\large The 22nd International Conference on Pervasive Computing and Communications (PerCom 2024)}}

\begin{document}
\title{{Stitching Satellites to the Edge: Pervasive and Efficient Federated LEO Satellite Learning}}

\author{\IEEEauthorblockN{
 Mohamed Elmahallawy, Tie Luo\IEEEauthorrefmark{1}}  
      \IEEEauthorblockA{%
Department of Computer Science, Missouri University of Science and Technology, Rolla, MO 65401, USA}
Emails:  \{meqxk, tluo\}@mst.edu
\thanks{\IEEEauthorrefmark{1}Corresponding author.\newline This work was supported by the National Science Foundation (NSF) under Grant No. 2008878.}  
}
\maketitle
\thispagestyle{mystyle}
\begin{abstract}
In the ambitious realm of space AI, the integration of federated learning (FL) with low Earth orbit (LEO) satellite constellations holds immense promise. However, many challenges persist in terms of feasibility, learning efficiency, and convergence. These hurdles stem from the bottleneck in communication, characterized by sporadic and irregular connectivity between LEO satellites and ground stations, coupled with the limited computation capability of satellite edge computing (SEC). This paper proposes a novel FL-SEC framework that empowers LEO satellites to execute large-scale machine learning (ML) tasks {\em onboard} efficiently. Its key components include i) {\em personalized learning via divide-and-conquer}, which identifies and eliminates redundant satellite images and converts complex multi-class classification problems to simple binary classification, enabling rapid and energy-efficient training of lightweight ML models suitable for IoT/edge devices on satellites; ii) {\em orbital model retraining}, which generates an aggregated ``orbital model'' per orbit and retrains it before sending to the ground station, significantly reducing the required communication rounds. We conducted experiments using Jetson Nano, an edge device closely mimicking the limited compute on LEO satellites, and a real satellite dataset. The results underscore the effectiveness of our approach, highlighting SEC's ability to run lightweight ML models on real and high-resolution satellite imagery. Our approach dramatically reduces FL convergence time by nearly 30 times, and satellite energy consumption down to as low as 1.38 watts, all while maintaining an exceptional accuracy of up to 96\%.
\end{abstract}


\section{Introduction}
The fusion of AI and edge computing with low Earth orbit (LEO) satellites holds promise for enhancing a multitude of space applications such as monitoring remote regions (deserts, forests, polar areas, etc.) and maritime zones for disaster management \cite{ perez2021airborne}. This potential is propelled by ongoing technological advancements in hardware capabilities and the burgeoning support for AI within the global community. Notably, rapid progress in camera technologies and the emergence of quantum computing are poised to usher in a new era of on-orbit AI \cite{abdelsadek2022future}. However, many challenges need to be addressed in this endeavor. For example, the irregular and intermittent visibility of LEO satellites to ground stations, constrained communication bandwidth, and data privacy and security, are substantial bottlenecks hindering the integration of AI into space applications. Prominent among these challenges is the severe limitation in the computing capacity of LEO satellites (e.g., CubSat and nanosatellite), which significantly impedes the training of large machine learning (ML) models.

Federated learning (FL) has emerged as a privacy-aware distributed ML approach \cite{mcmahan2017communication}. It enables clients (in this case LEO satellites) to collaboratively train an ML model onboard without transmitting raw data. This collaboration is orchestrated through a parameter server ($\mathcal{PS}$), which is typically a ground station (GS) in the case of LEO satellites. However, integrating FL with LEO satellites as edge nodes within the satellite edge computing (SEC) framework, presents significant challenges. These challenges arise from the inherent conflict between the restricted computing capabilities of LEO satellites and the vast volume of Earth observation images they collect, often necessary for training large-scale ML models \cite{wang2023satellite}. The process of training such models is power-intensive, demanding substantial computation, time, and energy---a heavy burden for LEO satellites. Consequently, the feasibility of conducting onboard ML for training global FL models across multiple orbits, as in an LEO constellation, becomes highly impractical.

On the other hand, addressing computing and power consumption concerns is not the sole obstacle to achieving rapid FL convergence and high efficiency within SEC. A critical factor lies in the inherent nature of FL, which relies on iterative communication rounds between clients and the ${\mathcal{PS}}$. This leads to a notably prolonged learning process within SEC due to the low satellite bandwidth and, more importantly, the highly intermittent and sporadic connections to the ${\mathcal{PS}}$. The latter arises from the substantial disparities between the trajectory of the Earth (rotating around the Axis) and that of LEO satellites (varying in any angle between 0 and 90 degrees w.r.t. the equator), as well as the large speed difference (29.78 km/s for the Earth and 7.8 km/s for satellites). Moreover, the visible windows of LEO satellites to the ${\mathcal{PS}}$ are typically short, lasting only a few minutes on average, rendering them inadequate for the transmission of large ML models. Consequently, these factors collectively extend the FL convergence time in SEC significantly, taking several days or even weeks \cite{chen2022satellite, so2022fedspace}.

In response to these challenges, this paper proposes a pragmatic FL framework tailored for SEC towards achieving pervasive space AI. This framework empowers LEO satellites, equipped with limited computing power, memory, and energy resources, to participate in collaborative onboard ML model training across extensive datasets spanning multiple orbits. Furthermore, it accelerates FL convergence significantly by substantially reducing the number of communication rounds while upholding competitive model accuracy. The following summarizes our main contributions.

\begin{itemize}[leftmargin=*]
\item We propose an scheme, {\em personalized learning via divide-and-conquer}. This scheme identifies and eliminates redundant satellite images on each individual satellite, effectively reducing data volume from billions to thousands without impairing model quality. Furthermore, it simplifies the original ML problem from a complex multi-class classification into a streamlined binary classification, enabling the training of lightweight ML models on each satellite. This approach is particularly advantageous for conserving energy and accommodating computing and memory limits on those small satellites. Notably, a subsequent model aggregation process restores the multi-class classification problem, thereby preserving the intended model quality and purpose.

\item We propose an {\em orbital model retraining} scheme, wherein a designated satellite within each orbit aggregates all models from satellites within the same orbit, forming an ``orbital model'', and subsequently, retrains it before transmitting to the $\mathcal{PS}$. This approach i) minimizes the waiting time for all satellites to individually enter their respective visible zones for model transmission to ${\mathcal {PS}}$, and ii) reduces the number of required communication rounds significantly, from typically tens to only a few. Ultimately, it leads to rapid global model convergence.

\item Using a real-world satellite imagery dataset for classification tasks, our experimental evaluations showcase remarkable efficiency. The proposed FL-SEC framework demonstrates swift convergence within a mere 2 hours while maintaining competitive accuracy levels. We measure accuracy using various metrics, including accuracy, precision, recall, and F1-score. Furthermore, our approach yields lightweight models with minimal computation, communication overhead, and energy consumption---significantly lower than previous solutions in the literature. These desired properties hold substantial promise, especially for resource-constrained LEO satellites that often rely on solar power.
\end{itemize}
 
\section{Related Work}

Despite the nascent stage of integrating FL into SEC, early attempts have been made \cite{chen2022satellite,razmi2022icc,happaper,elmahallawy2024communication,chen2023edge,e2023opt,elmahallawy2023secure,elmahallawy2023one,razmi2022ground,wang2022fl,mAsyFLEO,so2022fedspace,wu2023fedgsm}. For instance, some {\em synchronous FL approaches}, such as \cite{chen2022satellite}, employed the vanilla FedAvg \cite{mcmahan2017communication} with SEC without specific adaptations for communication challenges in LEO constellations. In the work of \cite{razmi2022icc}, FedISL was proposed, utilizing inter-satellite links (ISL) to expedite FL training in SEC. Another study, \cite{happaper,elmahallawy2024communication}, incorporated high altitude platforms (HAPs) like aerostats, stratospheric balloons, and stratollites as ${\mathcal{PS}}$s to expedite FL convergence. In \cite{chen2023edge}, the author proposed a clustering and edge selection approach wherein the GS forms clusters, each containing an LEO satellite as server and nearby LEO clients, based on channel quality. However, all these previous works simplified the problem by relying on simple datasets (e.g., MNIST \cite{deng2012mnist} and CIFAR-10 \cite{CIFAR-10}) for training tasks, and even then, they experienced long convergence times. This limitation significantly impedes their applicability in practical scenarios where each satellite trains an ML model using real satellite images, highlighting the need for more efficient and practical training approaches. Although the authors of \cite{e2023opt,elmahallawy2023secure} demonstrated the effectiveness of their approach on a realistic satellite dataset containing satellite images, they overlooked the necessary computing and communication resources required for model execution and transmission to the ${\mathcal{PS}}$. 

On the other hand, some {\em asynchronous FL approaches} have emerged as well. For instance, FedSat was introduced in \cite{razmi2022ground}, which involves averaging received satellite models in a regular order based on their visibility. It assumes regular satellites visit the GS, once per orbital period, aiming to ensure an equal contribution to the global model and to mitigate staleness issues typical in standard asynchronous FL approaches. In a similar vein, \cite{wang2022fl} introduced a graph-based routing and resource reservation algorithm to optimize FL model parameter transfer delays faced by \cite{razmi2022ground}. AsyncFLEO \cite{mAsyFLEO} tackled staleness by grouping models collected from satellites in different orbits based on their similarity, maintaining balance by selecting an equal number of models from each group. Fresh models are added without weighting, while stale ones are down-weighted based on their staleness. Another approach, known as FedSpace and proposed by So et al. \cite{so2022fedspace} attempted to strike a balance between the idleness associated with synchronous FL and the staleness issues associated with asynchronous FL. However, FedSpace relies on satellites uploading a portion of their raw data to the GS, which conflicts with the principles of efficient communication and data privacy in FL. Finally, Wu et al. \cite{wu2023fedgsm} propose FedGSM, a method addressing gradient staleness through a compensation mechanism. By harnessing the deterministic and time-varying orbit topology, FedGSM mitigates the adverse effects of staleness, yet it still exhibits slow convergence.

All these previous efforts to integrate FL with SEC have primarily aimed at addressing intermittent connections between LEO and the ${\mathcal{PS}}$ and expediting FL convergence. However, they commonly overlook the limited satellite computing resources and neglect power consumption optimization for on-edge model training, which limits their practicality in SEC scenarios.


\begin{table}[ht!]
\setlength{\tabcolsep}{0.4em}
\centering
\renewcommand{\arraystretch}{1.4}
\caption{Key notations for computation\&communication models.} 
\label{notation}
 \begin{tabular}{p{1cm} ||p{7cm} }
 \hline 
\multicolumn{2}{|c|}{\bf Computation Parameters}\\
\hline\hline
${\beta}$& Index to the global communication rounds\\
${V}$& Predefined number of orbital training rounds\\
$\boldsymbol{w}^{\beta}$& Global ML model\\
$\boldsymbol{w}_{i}^{\beta}$& Local model of a satellite $i$ in round $\beta$\\
$|\boldsymbol{w}|$&Size of the $\boldsymbol{w}$ in bits\\
$\eta$& Learning rate\\
$D_i$& Private data of a satellite $i$\\
$m_i$& Size of a SEC $i's$ data\\
$m^{\beta}$& Total data size of all satellites in round $\beta$\\
$\pi_{i}$& Filtering Policy for a satellite $i$'s collected Images\\
$C_{i}^{\text{CPU}}$&  Number of CPU cores of a satellite $i$\\
$f_{i}$&  CPU clock frequency of a satellite $i$ [GHz]\\
$t_{\text{filter}}^{i}$&Time require for a  satellite $i$ to filter its images [s]\\
$t_{\text{train}}^{i}$&Time require for a satellite $i$ to train an ML model [s]\\
\hline\hline
\multicolumn{2}{|c|}{\bf Communication Parameters}\\
\hline\hline
$P$& Transmitter power  (SEC or $\mathcal {PS}$) [dBm]\\
$G_{i/\mathcal{PS}}$& Antenna gain for a satellite $i$ or $\mathcal{PS}$ [dBi]\\
$T$& Noise temperature [K]\\
$K_B$& Boltzmann constant [J/K]\\
$B$& Bandwidth [GHz]\\
$\mathcal{L}_{i,\mathcal{PS}}$& Free-space path loss between a satellite $i$  and the $\mathcal{PS}$\\
$\mathnormal{d_{i,\mathcal{PS}}}$& Threshold distance for feasible LoS communication between a satellite $i$  and the $\mathcal{PS}$ [km]\\
$\lambda$& Signal wavelength [mm]\\
$R$& Achievable data rate between SEC  and the $\mathcal{PS}$ [Mb/s]\\
$t_{\text{wait}}$&Waiting time for a satellite $i$ to enter $\mathcal{PS}$'s visible zone [s]\\
$t_{\text{trans}}^{i,\mathcal{PS}}$& Transmission time of a model between satellite $i$ and $\mathcal{PS}$ [s]\\
$t_{\text{prop}}^{i,\mathcal{PS}}$&Propagation time for a model by a satellite $i$ or $\mathcal{PS}$ [s]\\

\hline\hline
\end{tabular}
\end{table}

\section{System model and Preliminaries}
For clarity, Table~\ref{notation} presents all the main notations used in this paper along with their corresponding descriptions.

\subsection{System Model}

    
We consider a generic LEO constellation in the context of Satellite Edge Computing (SEC). The constellation consists of $N$ orbits, each having a set $\mathcal{I}$ of equal-distanced satellites with unique IDs. Within each orbit, the satellites' computing and storage resources allocated to the FL task are denoted as $\mathcal{C} = \{c_1, c_2, \ldots, c_I\}$ and $\mathcal{S} = \{s_1, s_2, \ldots, s_I\}$, where $c_i \in [0, 1]$, $s_i \in [0, 1]$, and $I=|\mathcal{I}|$. Satellites in orbit $n$ travel around the Earth at an altitude $h_n$ with a speed of $v_{n}={\frac{2\pi{(R_{E}+h_{n}})}{T_{n}}}$, where $R_{E}$ is the radius of the Earth and $T_{n}$ is the orbital period given by $T_{n}= \frac{2 \pi}{\sqrt{G_EM}}{(R_{E}+h_{n})^{3/2}}$, with $G_E$ denoting the gravitational constant and $M$ representing the mass of the Earth. While orbiting the Earth, these satellites capture high-resolution Earth observation images that will be used for training ML models for a variety of classification tasks. This takes place under an FL framework, where each satellite initially receives a global model from the ${\mathcal{PS}}$ when it enters its respective visibility window. Subsequently, the satellite independently retrains the global model using its locally collected data and transmits the updated model back to the ${\mathcal{PS}}$, when its next visible window arrives. After collecting the model parameters from all the orbits, the ${\mathcal{PS}}$ aggregates them into an updated global model and distributes it once again to all satellites, initiating another communication round. This iterative process continues until the global model achieves convergence (i.e., meets a predefined termination criterion, such as reaching a target accuracy or loss, a maximum number of communication rounds, or negligible changes in the global model parameters).

\subsection{Preliminaries of Federated Learning in SEC}\label{sec:comm_para}

{\bf \underline{Computation Model.}}

In each communication round $\beta=0,1,\dots,\mathsf{B}$:
\begin{enumerate}[label=\arabic*),leftmargin=*]
    \item The ${\mathcal{PS}}$ transmits the global model $\boldsymbol{w}^{\beta}$ to each satellite when that satellite comes into its visible zone.
    \item Each satellite $i$ retrains $\boldsymbol {w}^{\beta}$ using a local optimization method such as stochastic gradient descent (SGD) on its collected Earth observation data, as
    \begin{equation}\label{eq:sgd}
    \boldsymbol {w}_{i}^{\beta,j+1} = \boldsymbol {w}_{i}^{\beta,j}- \eta \nabla F_{k}(\boldsymbol {w}_{i}^{\beta,j}; \{(X_{i}^{q}, y_{i}^{q})\})
    \end{equation}
    where $\boldsymbol {w}_{i}^{\beta,j}$ denotes $i$'s own copy of the model (referred to as a ``local model'') at the $j$-th local training epoch ($j=1,2,...,J$), $(X_{i}^{q}, y_{i}^{q})\subset D_{i}$ signifies the $q$-th mini-batch for each epoch $j$, $D_{i}$ is satellite $i$'s dataset, and $\eta$ denotes the learning rate. After completing the training, the satellite sends the updated local model $\boldsymbol {w}_{i}^{\beta,J}$ back to ${\mathcal{PS}}$.

    \item Upon receiving the updated local models from all satellites, the ${\mathcal{PS}}$ aggregates them into an updated global model, as $\boldsymbol {w}^{\beta+1} = \sum_{i\in \mathcal I} \frac{m_{i}}{m} \boldsymbol {w}_{i}^{\beta,J}$, where $m_i=|D_i|$ is the dataset size for satellite $i$ and $m=\sum_{i \in \mathcal I}  m_i$. The ${\mathcal{PS}}$ then distributes this updated global model to all satellites again, during their respective visibility windows, as described in (1).
  
\end{enumerate}

{\bf \underline{Communication Model.}} In an LEO constellation $\mathcal{I}$, a satellite $i$ and the $\mathcal{PS}$ 
can establish a line-of-sight (LoS) communication if the angle between their respective trajectories satisfies $\angle (r_{\mathcal{PS}}(t), (r_{i}(t) - r_{\mathcal{PS}}(t))) \leq \frac{\pi}{2}-\alpha_{min}$. Here, $r_{i}(t)$ and $r_{\mathcal{PS}}(t)$ are the trajectories of satellite $i$ and the $\mathcal{PS}$, respectively, and $\alpha_{min}$ is the minimum elevation angle. Furthermore, given an additive white Gaussian noise (AWGN) channel, the signal-to-noise ratio (SNR) between them can be expressed by
\begin{equation}
\text{SNR}_{i,\mathcal{PS}}=\frac{P G_{i}G_{\mathcal{PS}}}{T B K_B \mathcal{L}_{i,\mathcal{PS}}}
\end{equation}
where $P$ denotes the transmission power, $G_{i}$ and $G_{\mathcal{PS}}$ correspond to the antenna gains of satellite $i$ and $\mathcal{PS}$, respectively, $T$ stands for the noise temperature, $B$ is the channel bandwidth, $K_B$ denotes the Boltzmann constant, and $\mathcal{L}_{i,\mathcal{PS}}$ signifies the free-space path loss. This path loss can be determined as
\begin{equation}
    \mathcal{L}_{i,\mathcal{PS}} = \big(\frac{4\pi \|i,\mathcal{PS}\|_{2}}{\lambda}\big)^{2}
\end{equation}
under the condition that the Euclidean distance between satellite $i$ and $\mathcal{PS}$ satisfies $\|i,\mathcal{PS}\|_{2} \leq \mathnormal{d_{i,\mathcal{PS}}}$, where $\mathnormal{d_{i,\mathcal{PS}}}$ is the minimum distance that allows for LoS communication between satellite $i$ and $\mathcal{PS}$, and $\lambda$ is the wavelength of the transmitted signal. 

\begin{algorithm}
\caption{\fontsize{8.7}{0}\selectfont Personalized Learning via Divide-and-Conquer}\label{algorithm1}
\kwInit{Satellite constellation $\mathcal I$, all orbits $\mathcal{N}$, set of satellites $\mathcal I_n$ for each orbit $n\in\mathcal{N}$, satellite $i$'s captured data $D_i$, number of classes $L$. }
\kwInita{Set of binary-classifiers $\{\boldsymbol{w}_i^{\beta}\}_n$ for each orbit $n$.}
\ForEach(\\ \Comment*[h]{\scriptsize Divide Multi-classes Task into $L$ Binary Tasks}){$n\in \mathcal N$ }{
\If{$L\leq |\mathcal I_n|$}{
 Assign $L$ binary training tasks to $L$ satellites who have sufficient computing resources\\
 $\mathcal {I}_n= \{\text{the above $L$ satellites}\}$\\
 \ForEach{$i\in \mathcal {I}_n$}{
 $D_i^{\text{filtered}} \gets$ Filter $D_i$ for target class $l$ using \eqref{eq:filter}\\
 Train  a binary classifier model $\boldsymbol{w}_{i}^{\beta}$ using $D_i^{\text{filtered}}$\\
 $\{\boldsymbol{w}_i^{\beta}\}_n\gets\{\boldsymbol{w}_i^{\beta}\}_n\cup\boldsymbol{w}_{i}^{\beta}$ 
}
}
\Else{
Partition $L$ classes into $|\mathcal I_n|$ groups $\mathcal{G}$\\
\ForEach{$i\in \mathcal I_n$}{
 Assign  satellite $i$ with a group $G$ of binary tasks\\
 \ForEach{task $y\in G$}{
 \fontsize{8.7}{0}\selectfont $D_i^{\text{filtered}} \gets$ Filter $D_i$ for target class $y$ using \eqref{eq:filter}\\
 Train  a binary classifier model $\boldsymbol{w}_{i}^{\beta}$ on $D_i^{\text{filtered}}$\\
 $\{\boldsymbol{w}_i^{\beta}\}_n \gets\{\boldsymbol{w}_i^{\beta}\}_n\cup \boldsymbol w_{i}^{\beta}$ 
 } 
 }
}
Forward $\{\boldsymbol{w}_i^{\beta}\}_n$ w/ the filtered data size of each satellite $i$ to sink satellite for re-assembling (\sref{Sec:SinkSat})
}
\end{algorithm}


\section{Proposed Framework}\label{onorbit_model}


Recall that the hurdle of inefficiency in FL-SEC arises from several factors including the large number of communication rounds, sporadic and irregular visibility pattern of satellites, vast volume of data and limited computation and communication resources. These factors not only substantially slow down convergence but also give rise to notable temperature and power concerns \cite{ouyang2023joint}.

To tackle these challenges, we propose a two-fold approach: 1) In \sref{Sec:Sat_train}, we reduce the quantity of satellite images by filtering them based on their relevance to target classes, enabling each satellite to train a local model customized to a particular class associated with itself. This not only shrinks the model size to fit within the onboard processing capacity, but also involves resizing images to ensure they fall below the maximum allowable training time, all without sacrificing model accuracy. 2) In \sref{Sec:SinkSat}, we introduce a scheme where a designated satellite within each orbit generates and retrains an ``orbital model'' before transmitting it to the $\mathcal{PS}$. Not only does this overcome computation and energy constraints, but it also drastically reduces the number of required communication rounds. 

\subsection{Personalized Learning via Divide and Conquer}\label{Sec:Sat_train}

We take a divide-and-conquer (DnC) approach to divide the task of training a multi-class model into $L$ tasks, each involving the training of a binary classifier dedicated to a single class (where $L$ is the number of classes). If $L \leq I_n$ (i.e, number of satellites per orbit), each satellite will handle one such binary task; otherwise, multiple binary tasks are assigned and independently executed on each satellite. After completion of all such ``personalized'' training tasks, the trained binary models will be merged into the original intended multi-class classifier. The benefit of this method is that it substantially reduces each satellite's computation load, aligning well with the capability of SEC. 

\begin{figure}[t]
\centering
    { \includegraphics[width=\linewidth]{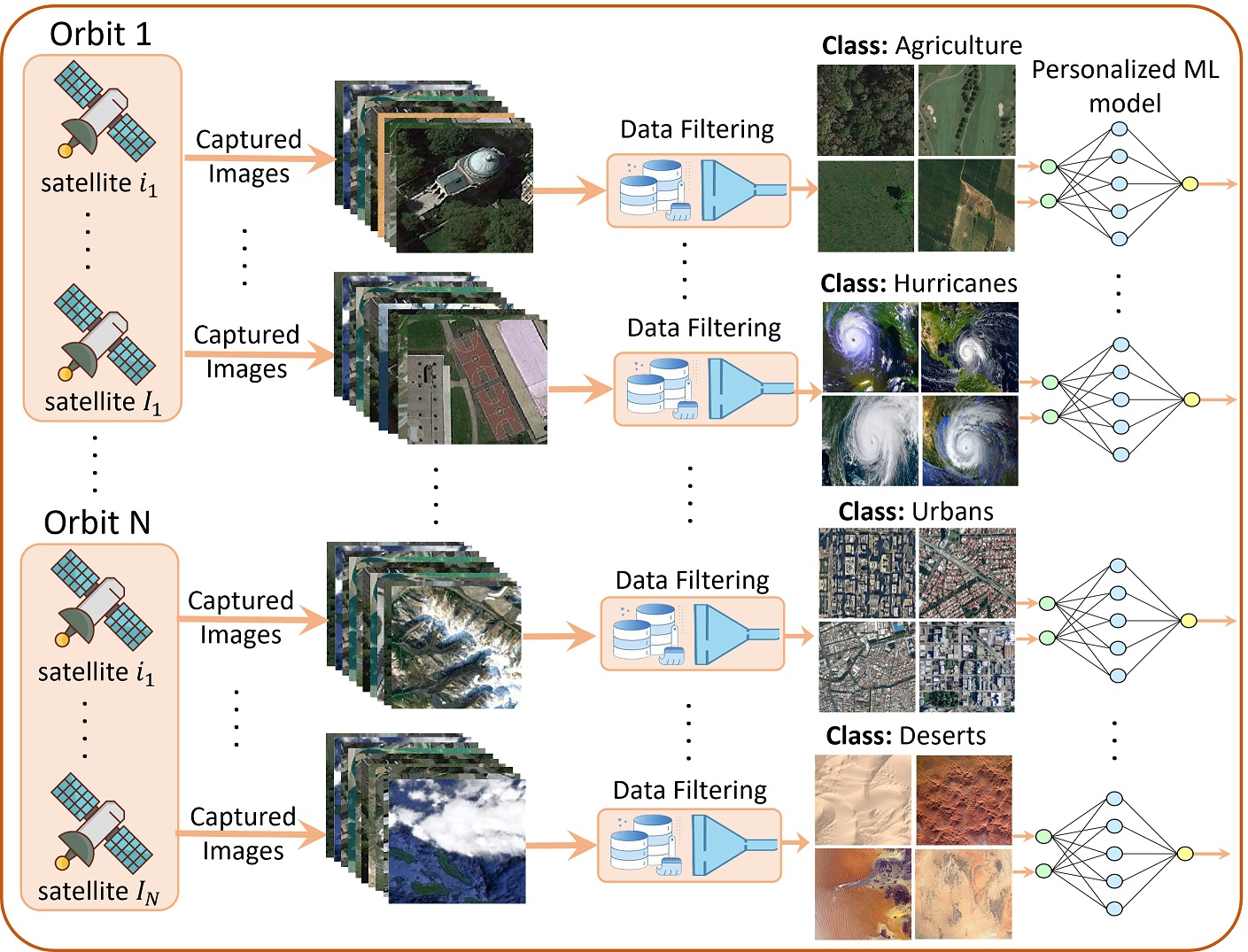}}
\caption{Personalized Learning via Divide-and-Conquer.}
\label{fig:data_filter}
\end{figure}

Algorithm~\ref{algorithm1} describes this approach which encompasses the following key elements. Each satellite utilizes only a subset of collected images, selected by their relevance to the satellite's target class while disregarding the rest. This strategic filtering reduces the number of training samples from millions to thousands, all while upholding the model's accuracy through an orbital retraining process described in \sref{Sec:SinkSat}. Additionally, each satellite resizes the chosen subset of images, further enhancing efficiency. These combined actions enable each satellite to complete its local model training within a matter of minutes, despite its limited computation capacity. Moreover, we formulate the minimum time required by each communication round while ensuring the gathering of satellite models from all orbits by the $\mathcal{PS}$. 
\fref{fig:data_filter} provides an overview.

We begin with the conventional FL-SEC approach, which uses a star topology as its communication architecture. Later in \sref{Sec:SinkSat}, we will replace it with a much more efficient architecture. In this conventional framework, we introduce a data filtering scheme. As such, the total time required for each satellite $i$ to receive the global model $\boldsymbol{w}^\beta$ from the $\mathcal{PS}$, filter its collected images, train $\boldsymbol{w}^\beta$ to obtain its local model $\boldsymbol{w}_i^\beta$, and transmit $\boldsymbol{w}_i^\beta$ to the $\mathcal{PS}$, can be formulated as
\begin{equation}\label{eq:sum}
t_{\text{req}}^{i} = t_{\text{wait}}^{{i}}+2 (t_{\text{trans}}^{i,\mathcal{PS}}+t_{\text{prop}}^{i,\mathcal{PS}})+t_{\text{filter}}^i+t_{\text{train}}^{i}
\end{equation}
where $t_{\text{wait}}^{{i}}$ is the waiting time for satellite $i$ to enter its visible zone, $t_{\text{trans}}^{i,\mathcal{PS}}=\frac{|\boldsymbol{w}|}{R}$ is the transmission time of a model between satellite $i$ and the $\mathcal{PS}$ (note that $\boldsymbol{w}^\beta$ and $\boldsymbol{w}_i^\beta$ have the same size, differing in parameter values only), $R=B~log_{_2}({1+SNR_{i, \mathcal{PS}}})$ is the achievable data rate between satellite $i$ and the $\mathcal{PS}$, $t_{\text{prop}}^{i,\mathcal{PS}}=\frac{\|i,\mathcal{PS}\|_2}{c}$ is the propagation delay between satellite $i$ and the $\mathcal {PS}$, $t_{\text{filter}}^i$ is the image filtering time, and $t_{\text{train}}$ is model train time. The filtering time can be expressed by
\begin{equation}\label{eq:time_filter}
t_{\text{filter}}^i =\frac{D_i~\pi_{i}}{C_i^{\text{CPU}}~f_{i}} 
\end{equation}
where $\pi_{i}$ is the policy employed by satellite $i$ to filter its collected images down to a threshold, 
$C_i^{\text{CPU}}$ is the number of CPU cores available on satellite $i$, and $f_{i}$ is its CPU clock frequency, which is adjustable through dynamic voltage and frequency scaling (DVFS) to minimize energy consumption. 

Our innovation toward efficient learning is to re-distribute the learning tasks among the satellites within the same orbit, such that each satellite is tasked with training a local model for classifying a single, specific class $y$. For instance, each class could pertain to a particular phenomenon such as a hurricane, forest fire, building footprint, flooding, or others. Hence, satellite $i$ utilizes a sampling policy $\pi_{i}$ to filter out images not belonging to its designated target class, thereby converting the original multi-class problem into one that can be solved by a binary classifier. These binary models from all satellites in orbit $n$ will later be re-assembled into a multi-class classifier in \sref{Sec:SinkSat} using a one-vs-all (OvA) strategy. The reduced dataset (after filtering), denoted by $D_i^{\text{filtered}}$, is given by
\begin{equation}\label{eq:filter}
D_i^{\text{filtered}} = \{ X_i \,|\, (X_i, y) \in D_{i}, \pi_i(y) = 1\}
\end{equation}
To identify images belonging to a target class, satellites have three options: (i) use the metadata of captured images, such as time, GPS location, satellite's attributes, etc. to determine whether an image is taken in the area of interest to that satellite that contains the target class of objects \cite{pritt2017satellite}; (ii) alternatively, satellites can store some labeled images beforehand or pre-train a model on them, and subsequently employ a semi-supervised classification method to label the newly collected data \cite{ostman2023decentralised}; (iii) satellites can also filter their data during the collection phase, where each satellite activates its camera on or off to only capture images of interest that belong to its target class \cite{denby2020orbital}. 
Thus, after our data reduction \eqref{eq:filter}, the learning task is converted from multi-class to binary classification, leading to lightweight models and lower satellite energy consumption. 
In Eq.~\eqref{eq:sum}, $t_{\text{train}}$ can be expressed by
\begin{equation}
     t_{\text{train}}^i=  \frac{ \nu_{i}\cdot J\cdot c_{\text{process}} +c_{\text{overhead}}}{f_{i}}
\end{equation}
where $\nu_{i}=\Bigl\lceil{\frac{m_i^{\text{filtered}}}{\kappa}}\Bigl\rceil$ is the total number of the mini-batches on satellite $i$ with $m_i^{\text{filtered}}=|D_i^{\text{filtered}}|$ denotes the number of selected images and $\kappa$ is the batch size, $J$ is the number of epochs as defined below \eqref{eq:sgd}, $c_{\text{process}}$ is the average number of processing cycles needed for training a mini-batch on a satellite, 
and $c_{\text{overhead}}$ is the CPU/GPU cycles required for processing any additional overheads. Here, implicitly, we assume homogeneous satellites and image sizes, which is justified by the fact that they all belong to the same constellation. Also note that our data filtering will lead to a much-reduced value for $\nu_{i}$, minimizing the training time.

Thus, the required time for completing all the local models within an orbit $n\in \mathcal N$ can be calculated by
\begin{align}\label{eqn:total}
t_{\text{req}}^{\mathcal {I}_{n}}=
\begin{cases}
\sum_{i\in\mathcal{I}_{n}} t_{\text{req}}^{i}, & \text{if~} t_{\text{req}}^{i}< t_{\text{visible}}^{i}\\\\ 
\sum_{i\in\mathcal{I}_{n}}  (t_{\text{req}}^{i}+\alpha~t_{\text{wait}}^{{i}}), & \text{if~} t_{\text{req}}^{i}\geq t_{\text{visible}}^{i}.
\end{cases}
\end{align}
where $\mathcal{I}_{n}$ is the set of all the satellites on orbit $n$, 
$\alpha$ denotes the number of revolutions (cycles) for the satellite to become visible to the $\mathcal{PS}$ for sending its trained model, and $t_{\text{visible}}^{i}$ is the length of the visible window. 
Thus, the minimum time required for the $\mathcal{PS}$ to collect all the satellite models, as in a global communication round, is
\begin{align}\label{eq:t_op}
t_{\text{req}}^{\mathcal{I}} = \max_{n\in \mathcal N} \{t_{\text{req}}^{\mathcal{I}_n}\}
\end{align}

Recall that this subsection assumes the conventional FL-SEC with a star topology. That approach follows a mostly sequential process where each satellite individually uploads its model to the $\mathcal{PS}$, requiring substantial time per communication round. In contrast, our proposed scheme needs only one satellite per orbit to communicate with $\mathcal{PS}$, 
as detailed in \sref{Sec:SinkSat}.

\subsection{Orbital Retraining and Convergence }\label{Sec:SinkSat}


The purpose of this scheme is to reconstruct the multi-class learning task by aggregating binary classifiers from all satellites within each orbit $n \in \mathcal{N}$ into an orbital model, using the OvA strategy. Subsequently, this orbital model is retrained through multiple orbital epochs until it achieves an acceptable level of accuracy before it is sent back to the $\mathcal{PS}$. This orbital training not only improves the accuracy of the orbital model, but also eliminates a large number of time-consuming global communication rounds by introducing the much faster orbit epochs. Algorithm~\ref{algorithm2} summarizes this approach.

\noindent {\bf Assumption 1.} Each satellite $i$ has sufficient training samples for its target class $y$. More formally, 
\[ F_{y}(\boldsymbol{w_i^J};D_{i}^{\text{filtered}})< \tau \]
where $F_{y}(\boldsymbol{w_i^J};D_{i}^{\text{filtered}})$ is the training loss of a satellite $i$ over the filtered dataset for a target class $y$, and $\tau$ is a loss threshold to control the convergence of the satellite's local model.

\begin{figure}[t]
    \centering
    \includegraphics[width=\linewidth]{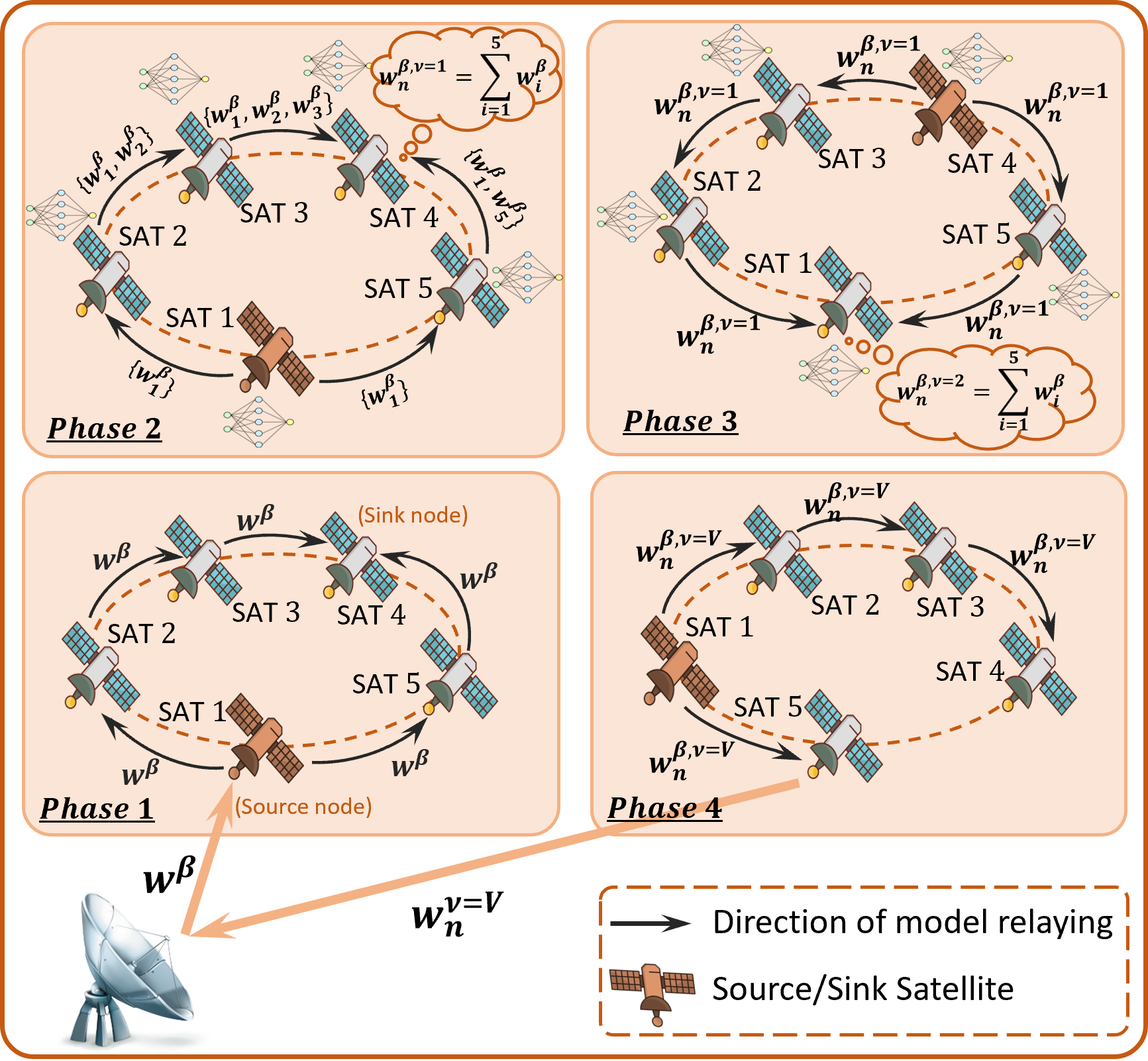}
        \caption{Illustration of Orbital Retraining (\sref{Sec:SinkSat}).}
\label{Framework}
\end{figure}

Our proposed scheme works as follows (cf. \fref{Framework}):
\begin{enumerate}[leftmargin=*,label=\arabic*)]
\item \underline{\bf \em Phase 1:} Within each orbit $n\in \mathcal{N}$, the first satellite who receives the global model $\boldsymbol {w}^{\beta}$ from the $\mathcal {PS}$, in this case \#1 and referred to as the ``source'' node, will initiate a model relaying process by forwarding $\boldsymbol {w}^{\beta}$ to its next-hop neighbors. The neighbors will then continue to relay $\boldsymbol {w}^{\beta}$ onward, and so forth until eventually reach a ``sink'' node. Each node starts to train $\boldsymbol {w}^{\beta}$ locally right after forwarding it (except that the sink node does not forward). Note that data filtering and resizing are also performed as per \sref{Sec:Sat_train} prior to training.

\item \underline{\bf \em Phase 2:} After training, each satellite $i$ obtains an updated $\boldsymbol {w}^{\beta}$ which is a local model ${\boldsymbol {w}}^{\beta}_{i}$ that satisfies Assumption 1. The source satellite, \#1, then initiates another round of relay by forwarding its trained model $\boldsymbol {w}_1^\beta$ to its next-hop neighbors, \#2 and \#5, who will then pass their own $\boldsymbol {w}_2^\beta$ and $\boldsymbol {w}_5^\beta$, together with the received $\boldsymbol {w}_1^\beta$, onto {\em their} respective neighbors ahead. This process continues until all satellites' models within that orbit reach the sink node \#4. This sink then aggregates all the received models into an ``orbital model'' $\boldsymbol {w}_n^{\beta,\upsilon}$, where the orbit index $n$ replaces satellite $i$, and $\upsilon=1,2,\dots,V$ indexes the intra-orbital training rounds. Trivially, any redundant models (in this case $\boldsymbol{w}_1^{\beta,\upsilon}$) will be removed before aggregation. Note that the orbital model {\em recovers} the intended multi-class classifier from the binary classifiers.  

\item \underline{\bf \em Phase 3:} After generating the orbital model $\boldsymbol {w}_n^{\beta,\upsilon=1}$, the sink satellite initiates a ``reverse relay'' and retraining process which is the same as Phases 1 and 2 except that the direction is from sink to source and the orbital model $\boldsymbol{w}_n^{\beta,\upsilon=1}$ replaces the global model $\boldsymbol{w}^{\beta}$. Subsequently, the source satellite \# will aggregate all the trained models $\boldsymbol {w}_{n,i}^{\beta,\upsilon=1}$ into an updated orbital model $\boldsymbol {w}_n^{\beta,\upsilon=2}$.

Thus, Phases 1-3 repeat (with the orbital model in place of the global model) to retrain the orbital model until $\upsilon=V$.

\item \underline{\bf \em Phase 4:} Finally, the satellite who aggregates $\boldsymbol{w}_n^{\beta,\upsilon=V}$, either source or sink, forwards $\boldsymbol{w}_n^{\beta,\upsilon=V}$ to all satellites on the same orbit (via relay) so that {\em any} satellite that becomes {\em visible first} can send $\boldsymbol{w}_n^{\beta,V}$ to the $\mathcal {PS}$ (for global aggregation among all orbits).
\end{enumerate}
\begin{algorithm}[!t]
\caption{ \fontsize{9.25}{0}\selectfont Orbital Retraining and Convergence Process}\label{algorithm2}
\kwInit{$\{\boldsymbol{w}_i^{\beta}\}_n$ for each orbit $n\in \mathcal{N}$, $m_i^\text{filtered}$ for all satellites, and $\beta$.}
\kwInita{Trained global model $\boldsymbol{w}^\mathsf{B}$.}
\ForEach{$\beta=0,1,\dots,\mathsf{B}$ }{
\ForEach(\\ \Comment*[h]{\fontsize{6.8}{0}\selectfont Convert Binary tasks back into a multi-class task}){$n\in\mathcal N$}{
Generate $\boldsymbol{w}_{n}^{\beta} = \sum_{i\in \mathcal I_{n}} \frac{m_{i}^{\text{filtered}}}{m_{\mathcal I_{n}}} \boldsymbol {w}_{i}^{\beta}$\\
Retrain $\boldsymbol{w}_{n}^{\beta}$ for $V$ orbital epochs as in \eqref{eq:Aggre}
}
Generate $\boldsymbol{w}^{\beta+1}=\sum_{n\in \mathcal{N}}\frac{|D_n|}{|D_{N}|}{\boldsymbol{w}}_{n}^{\beta,V}$\\
$\beta\gets \beta+1$
}

\end{algorithm}
Note that each model relay, in either a forward or reverse direction, is almost instantaneous because of the short inter-satellite distance and, optionally, the potential use of FSO in place of RF antenna. In contrast, in the model exchange between satellite and $\mathcal{PS}$, as in the conventional FL star topology, the delay is remarkably substantial because of the idle waiting for visible windows, as well as the long-distance (where only RF can be used).


This intra-orbit retraining process is important for two reasons: (1) it recovers the intended multi-class model from the binary classifiers ``personalized'' to each individual satellite; (2) the multi-round intra-orbit retraining takes maximal advantage of the feature learning capability from each personalized model and combines their best-learned patterns to effectively enhance the final aggregated model performance across all classes.

Below, we provide more details about Phases 1-4 outlined above. In \underline{\bf \em Phase 1}, the intra-orbit model relay eliminates the need for each node to wait for its visible window, thus ensuring each satellite to receive the global model at the earliest possible time. In \underline{\bf \em Phase 2}, the orbital model is aggregated as follows:
\begin{equation}\label{eq:Aggre}
   \Resize{7.9cm}{\boldsymbol{w}_{n}^{\beta,\upsilon+1} = \sum_{i\in \mathcal I_{n}} \frac{m_{i}^{\text{filtered}}}{m_{\mathcal I_{n}}^{}} \boldsymbol {w}_{i}^{J,\upsilon}, \text{\, \small where }  m_{\mathcal I_{n}}^{} = \sum_{i\in \mathcal I_{n}} m_{i}^{\text{filtered}}}
    \end{equation}
where $ \boldsymbol {w}_{n}^{\beta,\upsilon} $ is the orbital model for orbit $n$ at the $\upsilon$-th orbital iteration for a global round $\beta$. 
In \underline{\bf \em Phase 4}, when the visible satellite uploads the orbital model to the $\mathcal{PS}$, it will also upload {\em metadata} including the total data size $m_{\mathcal I_{n}}$ utilized for training $\boldsymbol{w}_{n}^{\beta,V}$, and the distribution of data (in terms of classes) across satellites within orbit $n$.

Corresponding to Eq.~\eqref{eq:sum} and \eqref{eq:t_op} which are formulated under the star topology, now we reformulate them under the new communication architecture involving model relay and retraining for $V$ iterations. Expressing the time needed for transmitting a model between two adjacent satellites via a single ISL hop as
\begin{equation}
t^{\text{ISL}}=\frac{h~|\boldsymbol{w}|} { B^{\text{ISL}} \varrho }
\end{equation}
where $h=1,2,\dots, H$ indexes the relay hops between source and sink satellites, $H=\lceil I_n/2 \rceil$ is the total hops due to the concurrent bilateral relaying in the same direction, $B^{\text{ISL}}$ denotes the allocated bandwidth for the ISL communication between two adjacent satellites, and $\varrho$ stands for the spectral efficiency of the communication link. Consequently, the overall model relay and retraining time on orbit $n$ is given by
\begin{align}\label{eq:orbit_time}
       t_{n}^{\text{ISL}} &= V \Big( \sum_{h=1}^{H}~t^{\text{ISL}}+ t_{\text{train}}\Big) + \\ \nonumber
       &\qquad t_{\text{filter}}+t_{\text{wait}}^{{n}}+ 2(t_{\text{trans}}^{n,\mathcal{PS}} + t_{\text{prop}}^{n,\mathcal{PS}})
\end{align}
Note that $t_{\text{wait}}^{{n}}$ is significantly smaller than $t_{\text{wait}}^{i}$ in Eq. \eqref{eq:sum}, because we only need the {\em first} node to become visible to transmit the orbital model to the $\mathcal{PS}$, while an arbitrary node as in \eqref{eq:sum} would wait for much longer. 

Now, we can express the minimum time required for the $\mathcal{PS}$ to obtain all the orbital models since the beginning of each round. This lower bound of time encompasses several components captured by \eqref{eq:t_op} and \eqref{eq:orbit_time}: (i) the waiting time for a satellite to enter its visible window, (ii) the transmission and propagation time between the $\mathcal{PS}$ and a visible satellite, either for the satellite to receive the global model $\boldsymbol{w}^\beta$ or for the $\mathcal{PS}$ to receive an orbital model $\boldsymbol {w}_{n}^{\beta,V}$, (iii) the time for data filtering and orbital model retraining within each orbit, and (iv) the time for intra-orbit model relay. As a result, Eq.~\eqref{eq:t_op} can be reformulated as
\begin{equation} \label{eq:req_time}
\begin{aligned}
     t_{\text{req}}^{{\mathcal I}} =  \max_{n\in\mathcal{N}} \bigl\{t_{n}^{\text{ISL}}\bigl\}    
\end{aligned}
\end{equation}
where $t_{n}^{\text{ISL}}$ as given by \eqref{eq:orbit_time} captures the above four time components.

Eqn.~\eqref{eq:req_time} reveals the benefit of orbital retraining which accelerates convergence by minimizing local model training time. In addition, it implies that the required time $t_{\text{req}}^{{\mathcal I}}$ for each round is controllable, 
by allowing each satellite to use only a subset of training images belonging to its target class, to reduce $t_{\text{train}}$ (since $t_{\text{wait}},t_{\text{trans}}~\text{and}~t_{\text{prop}}$ cannot be further reduced). On the other hand, this would compromise the global model accuracy. Therefore, a trade-off needs to be made between (i) crafting efficient ML models that align with satellite resources and (ii) achieving the desired convergence speed and accuracy. Via meticulous and empirical assignment of each satellite with reduced size of training data belonging to its target class, we have achieved significantly low computation and communication
overheads yet competitive accuracy as elaborated in \sref{sec:5} (see Table~\ref{train_cost}).

\section{Performance Evaluation}\label{sec:5}
\subsection{Experimental Setup}
\subsubsection{{LEO Constellation}}
We evaluate the convergence speed of our approach by employing two different configurations of the Walker Delta satellite constellation \cite{walker1984satellite}. These configurations differ in orbital inclination: 
one with an inclination angle of 45$^\circ$ (referred to as an {\em inclined constellation}) and the other with an inclination angle of 85$^\circ$ (referred to as {\em near-polar constellation}). Each constellation consists of 6 orbits, with each orbit $n$ containing ${I}_{n}=10$ equally spaced satellites. In both constellations, all satellites are positioned at an altitude  $h_n$ of 530 km. We consider a GS located in the central region of the U.S. (Rolla, Missouri) as our $\mathcal {PS}$, although it can be situated anywhere on the Earth. The GS has a minimum elevation angle of 10$^\circ$. To determine satellite-GS connectivity, we utilize a Systems Tool Kit simulator developed by AGI \cite{AGI}. We assign the computation 
 and communication parameters with their respective values as presented in Table \ref{parameter}. 

\begin{table}[!t]
\setlength{\tabcolsep}{0.3em}
\centering
\renewcommand{\arraystretch}{1.2}
\caption{Parameter Setting For our Performance Evaluation.} 
\label{parameter}

 \begin{tabular}{p{6.2cm}|| p{3cm}}
 \hline 
\multicolumn{2}{|c|}{\bf Computation}\\
\hline\hline
Number of orbital epochs (${V})$& \multicolumn{1}{c}{5}\\
Number of CPU cores of each satellite ($C_{i}^{\text{CPU}}$)& \multicolumn{1}{c}4\\
CPU clock frequency of each satellite ($f_{i}$)& \multicolumn{1}{c}{1.43 GHz} \\
\hline\hline
\multicolumn{2}{|c|}{\bf Communication}\\
\hline\hline
Transmitter power ($P$)& \multicolumn{1}{c}{60 dBm}\\
Antenna gain for any satellite or $\mathcal{PS}$ ($G_{i/\mathcal{PS}}$)& \multicolumn{1}{c}{6.98 dBi}\\
Noise temperature ($T$)& \multicolumn{1}{c}{354.81 K}\\
Bandwidth ($B$)& \multicolumn{1}{c}{0.5 GHz}\\
Signal wavelength ($\lambda$)& \multicolumn{1}{c}{15 mm}\\
Maximum data rate ($R$)& \multicolumn{1}{c}{16 Mb/s}\\
\hline\hline
\multicolumn{2}{|c|}{\bf SEC Training Models}\\
\hline\hline
Mini-batch size ($\kappa$)& \multicolumn{1}{c}4\\
Learning rate ($\eta$)&\multicolumn{1}{c}{0.001}\\
\hline\hline
\end{tabular}
\end{table}

\begin{figure}[!t]
\centering
    { \includegraphics[width=.75\linewidth]{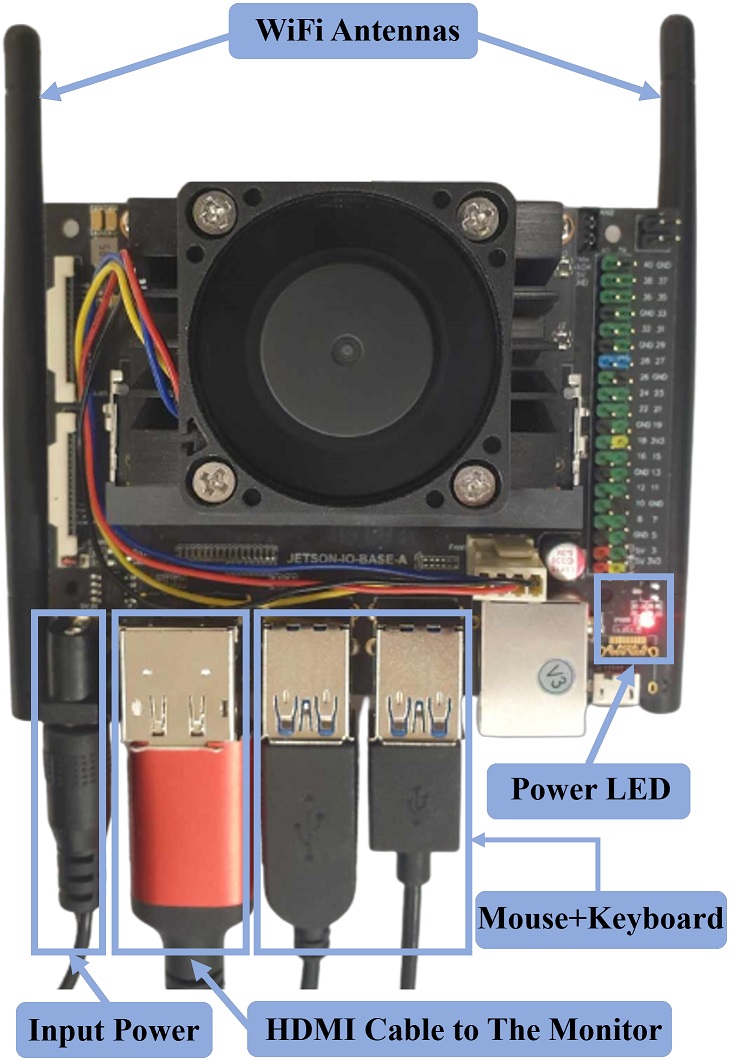}}

\caption{Our implementation, training, and testing on Jetson Nano with experimental setup of essential peripherals.}
\label{fig:Nano}
\end{figure}

\subsubsection{Implementation}
To ensure the compatibility of our approach with real satellite equipment, we implemented our approach on a Jetson Nano Dev-Kit\footnote{We are working with an Aerospace group in collaboration with NASA where an LEO satellite prototype has been built and the computing devices are indeed edge devices like Raspberry Pi and Jetson Nano.}. This energy-efficient device boasts a 128-core NVIDIA Maxwell GPU, a Quad-core ARM Cortex\circledR-A57 MPCore processor, 4GB of 64-bit memory, and runs on a remarkably low power consumption of just 5 watts  (can also be powered by a lithium-ion battery \cite{9944663}). During our experiments, we meticulously monitored both the computation overhead and the energy consumption associated with training tasks. Fig.~\ref{fig:Nano} shows our system setup.

\subsubsection{Datasets \& Satellites Training} 
To evaluate our approach, we train the satellite's local models using a real satellite dataset, EuroSat \cite{helber2019eurosat}. Additionally, we employ three other datasets—MNIST, CIFAR-10, and CIFAR-100—commonly used in state-of-the-art (SOTA) comparisons to assess our approach. Below, we provide a brief description of each dataset.

\begin{itemize}[leftmargin=*]
    \item  {\bf EuroSat \cite{helber2019eurosat}:} is a dataset that specializes in land use and land cover classification, tailored for remote sensing and Earth observation applications. This dataset encompasses high-resolution multispectral images captured by ESA's Sentinel-2 satellite mission. It consists of 27,000 satellite images, split into 21,600 training samples and 5,400 test samples. EuroSat delineates land into ten distinct classes, including agricultural land, industrial areas, residential areas, forests, rivers,  and others. Each image is precisely labeled with its corresponding land cover type and boasts a spatial resolution of 10 meters per pixel. For training our satellites, we utilize the VGG-16 model with trainable parameters of 8,413,194.
    
    \item \textbf{MNIST \cite{deng2012mnist}:}  is a dataset consisting of 70,000 grayscale images of handwritten numbers of size 28$\times$28 pixels. For training, we use a convolutional neural network (CNN) with three convolutional layers, three pooling layers, and one fully connected layer with 437,840 trainable parameters.

    \item \textbf{CIFAR-10 \& CIFAR-100 \cite{CIFAR-10}:} are two datasets, with the former containing 10 different classes and the latter containing 100 classes. Each dataset comprises 60,000 color images, evenly distributed across the classes. The images are all 32$\times$32 pixels in size and feature various animals and vehicles. For CIFAR-10, we use the same CNN architecture as MNIST, but with 798,653 trainable parameters. In the case of CIFAR-100, we use a CNN model constructed with six convolutional layers and two fully connected layers, totaling 7,759,521 training parameters.
\end{itemize} 
For the EuroSat, MNIST, and CIFAR-10 datasets, we evenly distributed the data size of each dataset across all orbits, with each orbit containing the 10 classes. Consequently, each satellite within each orbit was assigned only images of one class for training its local model. However, for the CIFAR-100 dataset, we assigned each satellite in each orbit 10 classes for training. We adopt this data distribution strategy to align with our goal of enabling each satellite to train a lightweight ML model. 

\subsection{Comparison with SOTA}

\begin{table}[!t]
\setlength{\tabcolsep}{0.7em}
\centering
\renewcommand{\arraystretch}{1.4}
\caption{Comparison of convergence time and accuracy under non-IID settings (near-polar Constellation).} 
\label{table_comp}
\resizebox{\linewidth}{!}{%
 \begin{tabular}{|p{0.2 cm}|p{2.1cm}|p{0.8cm} | p{1.19cm}| p{1.335cm}| p{4.37cm}|}
 \hline
 &\centering FL-LEO &\multicolumn{3}{c |} {Accuracy (\%)} &  Convergence time (h)\\
 \cline{3-5}
&\centering Approaches &\rmfamily MNIST& CIFAR-10& CIFAR-100& \\
 \hline \hline
\multirow{4}{*}&FedAvg \cite{mcmahan2017communication} & 79.41& 70.68& 61.66& 60 (\footnotesize$\mathcal{PS}$ located anywhere)\\
 \cline{2-6} 
&FedISL  \cite{razmi2022icc} &  82.76& 73.62& 66.57  & 8 (\footnotesize$\mathcal{PS}$ located at the NP)\\
 \cline{2-6} 
&FedISL \cite{razmi2022icc} &  61.06& 52.11& 47.99&72 (\footnotesize$\mathcal{PS}$ located anywhere)\\
  \cline{2-6} 
\multirow{-4}{2mm}{\rotatebox[origin=p]{90}{\centering {\small{Sync FL}}}}&NomaFedHAP \cite{elmahallawy2024communication} &  82.73& 77.36&  62.81&  24 (\footnotesize$\mathcal{PS}$ located anywhere)\\
 \hline    \hline
 \multirow{3}{*}&{FedAsync} \cite{xie2019asynchronous} & 70.36& 61.81& 56.37& 48 (\footnotesize$\mathcal{PS}$ located anywhere)\\
 \cline{2-6} 
 &FedSpace \cite{so2022fedspace}& 52.67&39.41& 36.04&72 (\scriptsize satellite uploads some of its data) \\ 
\cline{2-6} 
\multirow{-3}{3mm}{\rotatebox[origin=p]{90}{\centering {\small{Async FL}}}}&AsyncFLEO \cite{mAsyFLEO}&  79.49& 69.88& 61.43&9 (\scriptsize  sink satellite has sufficient visible period)\\
 \hline \hline   \rowcolor{red!20} 
 &  \textbf{Ours} &  \textbf{94.64}& \textbf{89.69}& \textbf{82.65}& \textbf{2.13} ($\mathcal{PS}$ located anywhere)\\
 \hline 
\end{tabular}}
\end{table}

We compare our approach with six other baseline methods: three synchronous FL methods, namely FedAvg \cite{mcmahan2017communication}, FedISL \cite{razmi2022icc}, and NomaFedHAP \cite{elmahallawy2024communication}, as well as three asynchronous FL methods, including FedAsync \cite{xie2019asynchronous}, FedSpace \cite{so2022fedspace}, and AsyncFLEO \cite{mAsyFLEO}. However, our approach is the first to account for computation and energy consumption overheads, ensuring that each satellite can run the FL approach with its limited computational and storage resources.

Compared to the baselines, our approach achieves convergence in approximately 2 hours with accuracies of 94.64\%, 89.69\%, and 82.65\% on MNIST, CIFAR-10, and CIFAR-100, respectively, outperforming all the baseline methods. The second-fastest approach is FedISL \cite{razmi2022icc}, synchronous FL approach, that attains accuracies of 82.76\%, 73.62\%, and 66.57\% on MNIST, CIFAR-10, and CIFAR-100, respectively, after 8 hours. When compared to asynchronous FL approaches, AsyncFLEO is considered the third-fastest method to converge, achieving accuracies of 79.49\%, 69.88\%, and 61.43\% on MNIST, CIFAR-10, and CIFAR-100, respectively. These results demonstrate the effectiveness of our approach, not only in making satellites applicable to realistic FL scenarios but also in achieving faster convergence with high accuracy. Table~\ref{table_comp} summarizes the comparison of our approach with the rest of the baseline approaches.


\subsection{In-Depth Evaluation on Eurostat Dataset}




\subsubsection{Convergence Results}
To evaluate the performance of our approach on the EuroSAT dataset, we utilize four evaluation metrics: accuracy, precision, recall, and F1-score. We first evaluate the convergence of our proposed approach under the near-polar constellation scenario for {\em only two} global communication rounds (because our approach can converge within very few rounds). After the first round, our approach achieved an average accuracy of 95.741\%. This first round required 124.86 minutes of communication time to ensure that at least one satellite per orbit received the global model, allowing each orbit to initiate the orbital training process. Furthermore, the training of orbital models was completed within a maximum training time of 27.0371 minutes across all orbits, resulting in an overall communication and training time of 151.891 minutes ($\sim$2.53 hours).
\begin{figure}[!t]
\centering
    \includegraphics[width=1\linewidth]{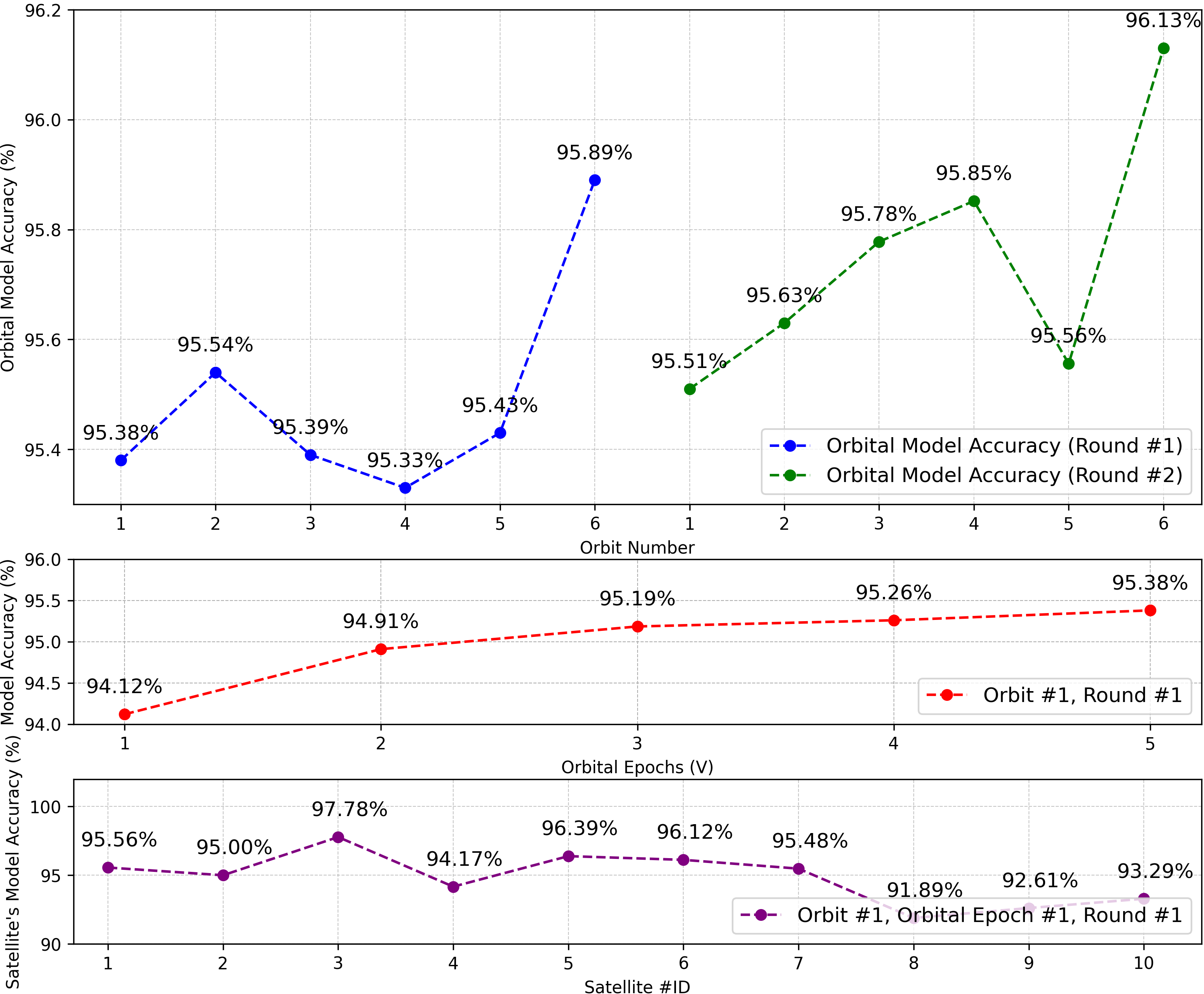}
\caption{Orbital model accuracy comparison across global communication rounds and orbital epochs. The upper subplot shows the accuracy of orbital models over two global communication rounds. The middle subplot illustrates the evolution of orbital model accuracy for orbit \#1 in global round \#1 over 5 orbital epochs. The lower subplot shows the training accuracy for each satellite model on orbit \#1 in orbital epoch \#1 and global round \#1. The high accuracy is attributed to our DnC approach that converts complex tasks into binary tasks.}
\label{fig:orbial_model}
\end{figure}
\begin{table}[!t]
\setlength{\tabcolsep}{0.55em}
\centering
\renewcommand{\arraystretch}{1.6}
\caption{{Evaluation of our approach on EuroSat Dataset.}} 
\label{EuroSat_eval}
\resizebox{\linewidth}{!}{%

 \begin{tabular}{|p{2.1cm}|p{0.7cm} ||p{0.86cm}|p{0.83cm}| p{0.865cm}| p{0.82cm}||p{0.86cm}|p{0.83cm} |p{0.865cm}| p{0.82cm}||}
 \hline  
\multirow{2}{*}{\diagbox[width=2.4cm]{Class}{\rotatebox{0}{Metric \hskip 0em}}} &\# of  &\multicolumn{4}{c||} {Near-polar constellation (85$^\circ$)}&\multicolumn{4}{c||} {Inclined constellation (45$^\circ$)}\\
\cline{3-10}
&images & ACC(\%)&PC (\%)& RC (\%)& F1  (\%)& ACC(\%)&PC (\%)& RC (\%)& F1(\%) \\
\hline \hline 
 AnnualCrop &600&97.63&92.06&94.67&93.34& 98.39&91.71&94.0&92.84\\
\hline
Forest &600&98.19&96.73&98.67&97.69&99.52&96.59&99.17&97.86\\
\hline
\scriptsize{HerbaceousVegetation}&600&99.32&93.05&93.67&93.36&98.5&92.12&94.5&93.33\\
\hline
 Highway &500&99.61&97.74&95.0&96.35&99.31&97.93&94.60&96.24\\
\hline
 Industrial &500&98.89&98.95&93.80&96.30&99.30&98.94&93.40&96.09\\
\hline
 Pasture &400&99.23&96.42&94.25&95.32&99.33&96.67&94.25&95.44\\
\hline
 PermanentCrop &500&97.46&95.14&90.0&92.50&98.54&94.50&89.40&91.88\\
\hline
 Residential &600&99.01&91.45&99.83&95.46&99.04&92.15&99.83&95.84\\
\hline
 River &500&99.56&98.17&96.60&97.38&99.44&97.37&96.60&96.99\\
\hline
 SeaLake &600&99.11&98.66&98.0&98.33&99.63&99.15&97.50&98.32\\

\hline\hline
\end{tabular}}
\end{table}

After the second round, our approach's accuracy showed a marginal increase to 95.778\%, with the total time for both rounds amounting to 277.167 minutes ($\sim$4.619 hours). The upper subplot of Fig.~\ref{fig:orbial_model} shows the accuracy of the orbital model for each orbit after 5 orbital epochs for two communication rounds. The middle subplot of Fig.~\ref{fig:orbial_model} illustrates the improvement of retraining the orbital model accuracy for a single orbit for each orbital epoch within the first global round. In addition, the lower subplot of Fig.~\ref{fig:orbial_model} displays the individual training accuracy of each satellite that trains a binary-class classifier before aggregating them into an orbital model. After averaging them and testing the orbital model's accuracy, it yields an accuracy of 94.12\%.
These results underscore the effectiveness of our orbital retraining approach in significantly reducing the required number of communication rounds to just a few iterations and expediting convergence.
\begin{figure}[!t]
\centering
 \subfloat[Near-polar constellation (85$^\circ$).]
 { \includegraphics[width=0.5\linewidth]{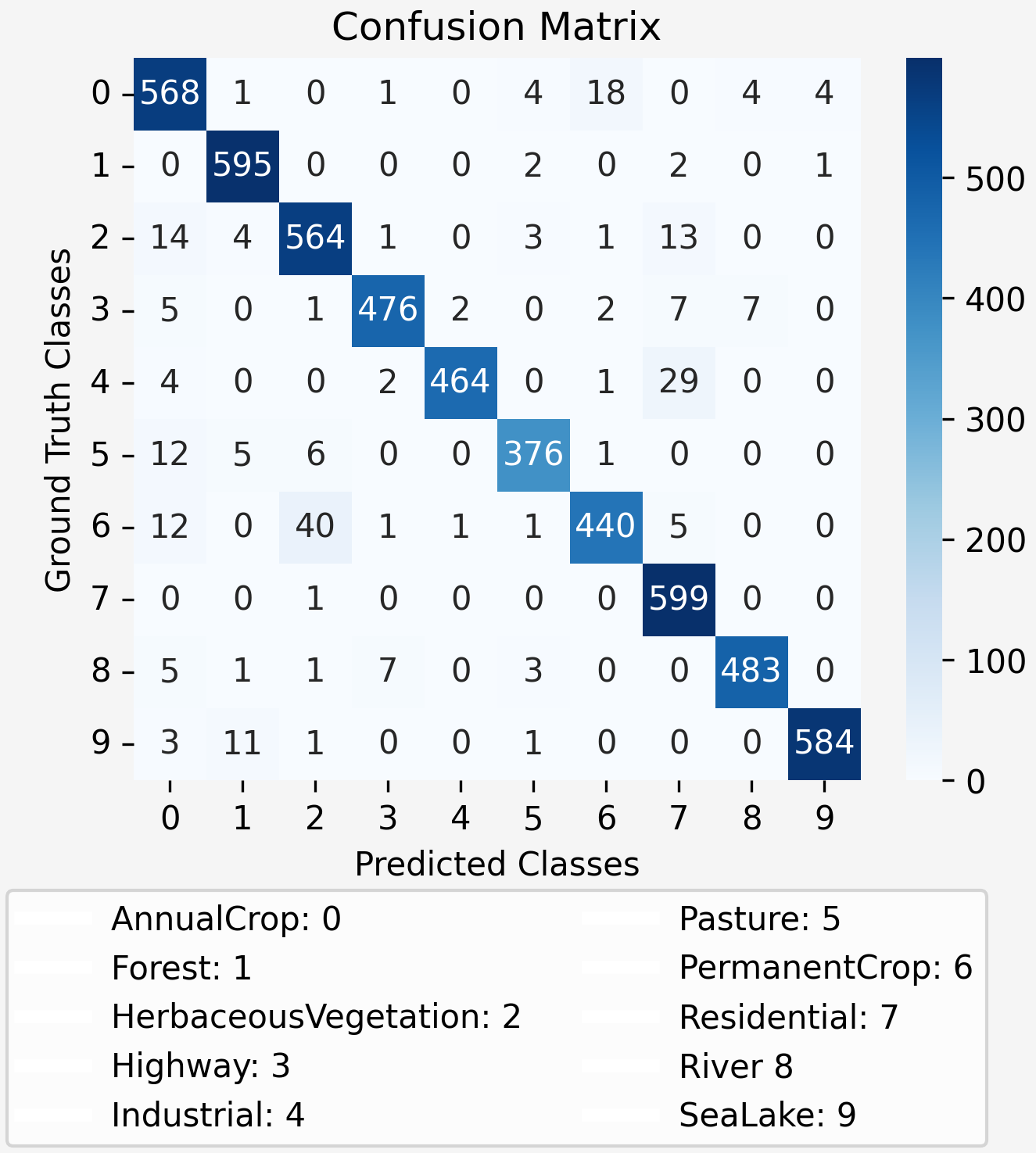 }}
 \subfloat[Inclined constellation (45$^\circ$).]
 { \includegraphics[width=0.5\linewidth]{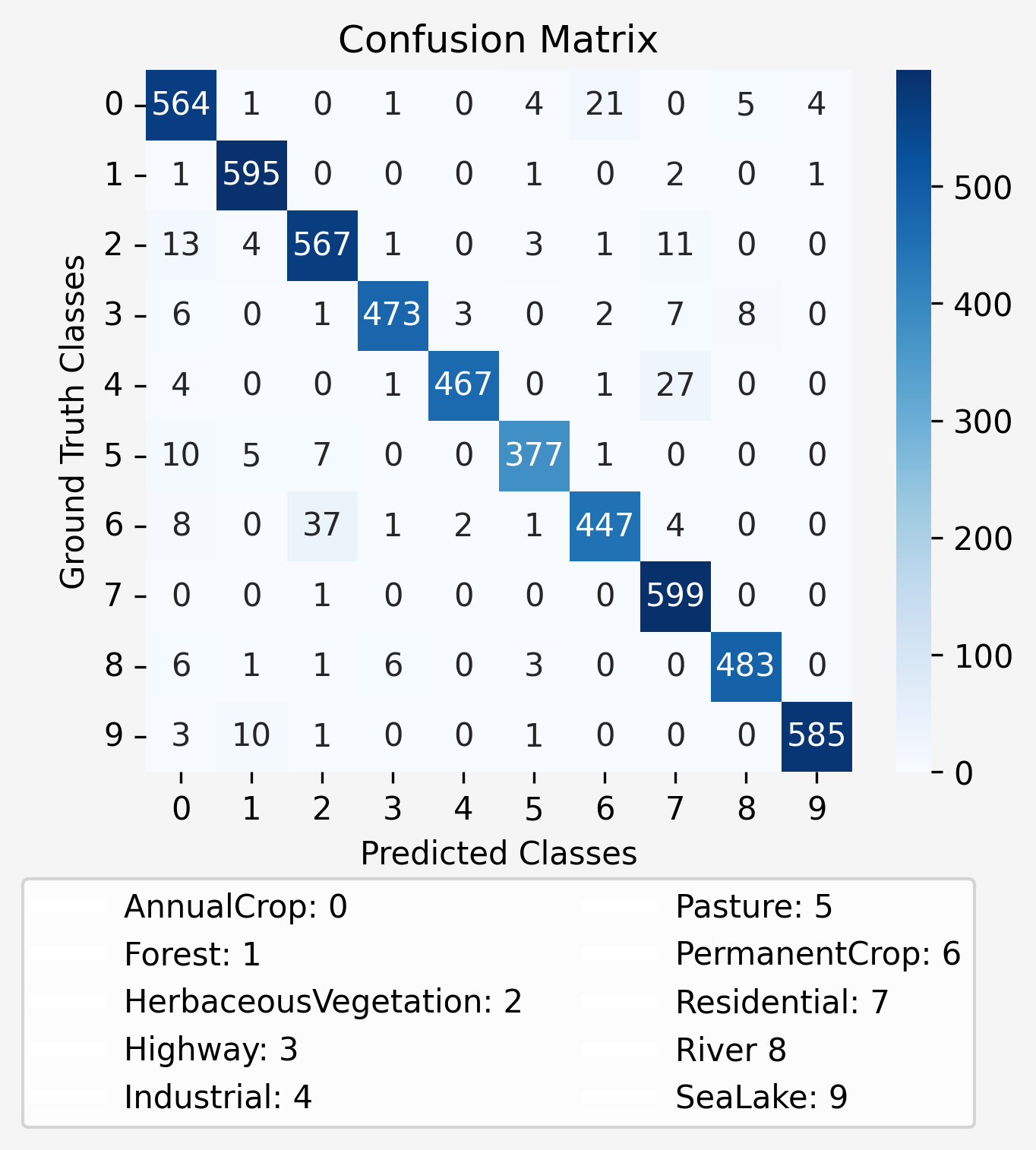 }}
\caption{Confusion matrix that compares 10 predicted and ground-truth classes for 5400 test images.}
\label{fig:conf_matrix}
\end{figure}

\begin{figure}[!t]
\centering
    \includegraphics[width=1\linewidth]{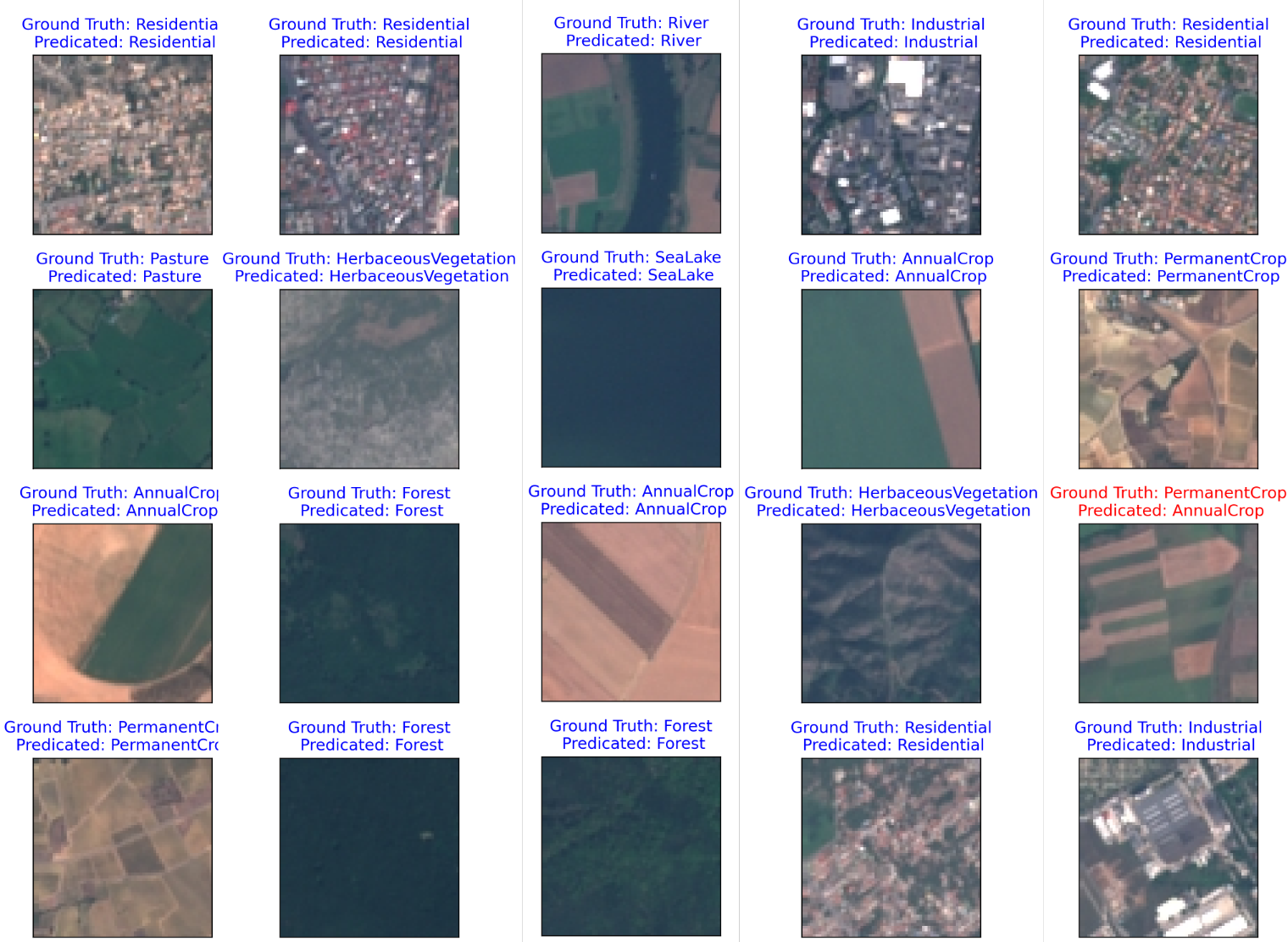}
\caption{Twenty randomly selected images from a Eurostat test set of 5400 samples, illustrating the predicted vs. ground truth labels. Blue and red color represent correct and incorrect predictions, respectively.}
\label{fig:Eurosat_eval}
\end{figure}

Table~\ref{EuroSat_eval} summarizes our approach's evaluation results under various metrics for each class after only two global training rounds. Furthermore, in Fig.~\ref{fig:conf_matrix}, we present the confusion matrix, which visually illustrates the relationship between actual ground truth classes and predicted classes, providing insights into the efficacy of our approach. Finally, in Fig.~\ref{fig:Eurosat_eval}, we showcase 20 randomly selected images from a pool of 5400 test samples to demonstrate the effectiveness of our approach. Remarkably, among these randomly selected images, only one is misclassified. This outcome further underscores the effectiveness of our approach in accurately classifying images. 

\begin{table}[!t]
  \caption{Comparison of computation and communication overheads of our approach in different settings.}
\label{train_cost}
  \begin{subtable}{0.56\linewidth}
    \centering
    \caption{\fontsize{8.3}{0}\selectfont Computation overhead.}
    \begin{tabular}{c|c}
      \toprule
      Model & FLOPS (G) \\
      \midrule
      CNN using MNIST  &  11.91\\
      CNN using CIFAR-10  & 15.58 \\
      CNN using CIFAR-100& 28.13 \\
      VGG-16 using EuroSat  & 43.84 \\      
      \bottomrule
    \end{tabular}
  \end{subtable}
  \begin{subtable}{.42\linewidth}
    \centering
        \caption{\fontsize{8.3}{0}\selectfont Communication overhead.}
    \begin{tabular}{c|c}
      \toprule
      Dataset & Size (MB) \\
      \midrule
      MNIST & 0.437 \\
      CIFAR-10& 0.798 \\
      CIFAR-100& 7.76 \\
      EuroSat & 26.68 \\
      \bottomrule
    \end{tabular}
  \end{subtable}%

\end{table}

\subsubsection{Overheads and Energy Consumption Results}
We analyze the required computation and communication overheads for running our proposed approach on edge devices, specifically the Jetson-Nano, to determine its affordability for deployment on satellites with limited resources. Our findings reveal that GPU usage during training ranges from 17\% to 58\%, with corresponding energy consumption levels ranging between 1.38 and 2.25 watts for each local model, demonstrating the model's lightweight nature.

In Table~\ref{train_cost}, we present both the computation and communication overheads of our approach for various models trained on different datasets, where each satellite is assigned a reduced data size belonging to its target class. 
To assess computation overheads, we use floating-point operations per second (FLOPS) as a metric to measure the computation cost. 
In Table~\ref{train_cost}.a, we can observe that the highest computation cost incurred by our approach is 43.84 GFLOPS. This is significantly lower than the computational capacity of the Jetson-Nano, which can handle up to 472 GFLOPS.


On the other hand, we evaluate the communication overhead of our approach in terms of the orbital model parameters that need to be uploaded to the $\mathcal{PS}$. In Table~\ref{train_cost}.b, we present the communication cost for all the training datasets used to evaluate our approach, with a maximum of 26.68 MB. Considering that a satellite has a short visible window of 5 minutes and needs to upload the orbital model of size 26.68 MB, this requires a small data rate of 0.69 Mb/s, which is significantly lower than the available data rate in Satcom. In summary, our approach exhibits efficient computation and communication overheads, making it suitable for space-edge computing scenarios and pervasive computing.

\section{Conclusion} \label{Conclusion} 
This paper represents a significant stride towards enabling pervasive space AI by seamlessly integrating FL with SEC in the form of LEO satellites. Our innovative approach comprises (1) {\em personalized learning via DnC}, which reduces the complexity of learning at SEC by transforming multi-class training into binary training, and (2) {\em orbital model retraining}, which aggregates and retrains an orbital model in multiple orbital epochs in a back-and-forth manner among all the satellites within an orbit, before sending it to the $\mathcal {PS}$. Our experiments with a real satellite imagery dataset validate the effectiveness of our approach, converging rapidly within only 4.619 hours while achieving a high classification accuracy of 95.778\%. Notably, it significantly reduces energy consumption for training local models at each satellite to as low as 1.38 watts, computation overhead to 11.91 GFLOPs, and communication overhead to 0.437 MB. These results demonstrate the suitability and applicability of our approach in real SEC scenarios with constrained resources.



{\small
\bibliographystyle{IEEEtran}
\bibliography{biblio.bib}}

\begin{thebibliography}{10}
\providecommand{\url}[1]{#1}
\csname url@samestyle\endcsname
\providecommand{\newblock}{\relax}
\providecommand{\bibinfo}[2]{#2}
\providecommand{\BIBentrySTDinterwordspacing}{\spaceskip=0pt\relax}
\providecommand{\BIBentryALTinterwordstretchfactor}{4}
\providecommand{\BIBentryALTinterwordspacing}{\spaceskip=\fontdimen2\font plus
\BIBentryALTinterwordstretchfactor\fontdimen3\font minus \fontdimen4\font\relax}
\providecommand{\BIBforeignlanguage}[2]{{%
\expandafter\ifx\csname l@#1\endcsname\relax
\typeout{** WARNING: IEEEtran.bst: No hyphenation pattern has been}%
\typeout{** loaded for the language `#1'. Using the pattern for}%
\typeout{** the default language instead.}%
\else
\language=\csname l@#1\endcsname
\fi
#2}}
\providecommand{\BIBdecl}{\relax}
\BIBdecl

\bibitem{perez2021airborne}
A.~Perez-Portero, J.~F. Munoz-Martin, H.~Park, and A.~Camps, ``Airborne gnss-r: A key enabling technology for environmental monitoring,'' \emph{IEEE Jnl. of Sel. Topics in Applied Earth Observations and Remote Sensing}, vol.~14, pp. 6652--6661, 2021.

\bibitem{abdelsadek2022future}
M.~Y. Abdelsadek, A.~U. Chaudhry, T.~Darwish, E.~Erdogan, G.~Karabulut-Kurt, P.~G. Madoery, O.~B. Yahia, and H.~Yanikomeroglu, ``Future space networks: Toward the next giant leap for humankind,'' \emph{IEEE Transactions on Communications}, vol.~71, no.~2, pp. 949--1007, 2022.

\bibitem{mcmahan2017communication}
B.~McMahan, E.~Moore, D.~Ramage, S.~Hampson, and B.~A. y~Arcas, ``Communication-efficient learning of deep networks from decentralized data,'' in \emph{AISTATS}, 2017, pp. 1273--1282.

\bibitem{wang2023satellite}
S.~Wang and Q.~Li, ``Satellite computing: Vision and challenges,'' \emph{IEEE Internet of Things Journal}, 2023.

\bibitem{chen2022satellite}
H.~Chen, M.~Xiao, and Z.~Pang, ``Satellite-based computing networks with federated learning,'' \emph{IEEE Wireless Communications}, vol.~29, no.~1, pp. 78--84, 2022.

\bibitem{so2022fedspace}
J.~So \emph{et~al.}, ``Fedspace: An efficient federated learning framework at satellites and ground stations,'' \emph{arXiv preprint arXiv:2202.01267}, 2022.

\bibitem{razmi2022icc}
N.~Razmi \emph{et~al.}, ``On-board federated learning for dense {LEO} constellations,'' in \emph{IEEE ICC}, May 2022.

\bibitem{happaper}
M.~Elmahallawy and T.~Luo, ``{FedHAP}: Fast federated learning for {LEO} constellations using collaborative {HAPs},'' in \emph{2022 14th International Conference on Wireless Communications and Signal Processing (WCSP)}, 2022, pp. 888--893.

\bibitem{elmahallawy2024communication}
M.~Elmahallawy, T.~Luo, and K.~Ramadan, ``Communication-efficient federated learning for {LEO} satellite networks integrated with {HAPs} using hybrid {NOMA-OFDM},'' \emph{IEEE Journal on Selected Areas in Communications}, pp. 1--1, 2024.

\bibitem{chen2023edge}
C.-Y. Chen, L.-H. Shen, K.-T. Feng, L.-L. Yang, and J.-M. Wu, ``Edge selection and clustering for federated learning in optical inter-{LEO} satellite constellation,'' \emph{arXiv preprint arXiv:2303.16071}, 2023.

\bibitem{e2023opt}
M.~Elmahallawy and T.~Luo, ``Optimizing federated learning in {LEO} satellite constellations via intra-plane model propagation and sink satellite scheduling,'' in \emph{ICC 2023 - IEEE International Conference on Communications}, 2023, pp. 3444--3449.

\bibitem{elmahallawy2023secure}
M.~Elmahallawy, T.~Luo, and M.~I. Ibrahem, ``Secure and efficient federated learning in {LEO} constellations using decentralized key generation and on-orbit model aggregation,'' in \emph{GLOBECOM 2023-2023 IEEE Global Communications Conference}, 2023.

\bibitem{elmahallawy2023one}
M.~Elmahallawy and T.~Luo, ``One-shot federated learning for {LEO} constellations that reduces convergence time from days to 90 minutes,'' in \emph{2023 24th IEEE International Conference on Mobile Data Management (MDM)}, 2023, pp. 45--54.

\bibitem{razmi2022ground}
N.~Razmi, B.~Matthiesen, A.~Dekorsy, and P.~Popovski, ``Ground-assisted federated learning in {LEO} satellite constellations,'' \emph{IEEE Wireless Communications Letters}, 2022.

\bibitem{wang2022fl}
P.~Wang, H.~Li, and B.~Chen, ``{FL}-task-aware routing and resource reservation over satellite networks,'' in \emph{GLOBECOM 2022-2022 IEEE Global Communications Conference}.\hskip 1em plus 0.5em minus 0.4em\relax IEEE, 2022, pp. 2382--2387.

\bibitem{mAsyFLEO}
M.~Elmahallawy and T.~Luo, ``{AsyncFLEO}: {A}synchronous federated learning for {LEO} satellite constellations with high-altitude platforms,'' in \emph{2022 IEEE International Conference on Big Data (BigData)}, 2022, pp. 5478--5487.

\bibitem{wu2023fedgsm}
L.~Wu and J.~Zhang, ``{FedGSM:} {E}fficient federated learning for {LEO} constellations with gradient staleness mitigation,'' \emph{arXiv preprint arXiv:2304.08537}, 2023.

\bibitem{deng2012mnist}
L.~Deng, ``The {MNIST} database of handwritten digit images for machine learning research,'' \emph{IEEE Signal Processing Magazine}, vol.~29, no.~6, pp. 141--142, 2012.

\bibitem{CIFAR-10}
A.~Krizhevsky, V.~Nair, and G.~Hinton, ``{CIFAR-10} (canadian institute for advanced research),'' \emph{URL http://www. cs. toronto. edu/kriz/cifar. html}, vol.~5, no.~4, p.~1, 2010.

\bibitem{ouyang2023joint}
Q.~Ouyang, N.~Ye, J.~Gao, A.~Wang, and L.~Zhao, ``Joint in-orbit computation and communication for minimizing download time from {LEO} satellites,'' \emph{IEEE Transactions on Mobile Computing}, 2023.

\bibitem{pritt2017satellite}
M.~Pritt and G.~Chern, ``Satellite image classification with deep learning,'' in \emph{2017 IEEE applied imagery pattern recognition workshop (AIPR)}.\hskip 1em plus 0.5em minus 0.4em\relax IEEE, 2017, pp. 1--7.

\bibitem{ostman2023decentralised}
J.~{\"O}stman, P.~Gomez, V.~M. Shreenath, and G.~Meoni, ``Decentralised semi-supervised onboard learning for scene classification in low-earth orbit,'' \emph{arXiv preprint arXiv:2305.04059}, 2023.

\bibitem{denby2020orbital}
B.~Denby and B.~Lucia, ``Orbital edge computing: Nanosatellite constellations as a new class of computer system,'' in \emph{Proceedings of the Twenty-Fifth International Conference on Architectural Support for Programming Languages and Operating Systems}, 2020, pp. 939--954.

\bibitem{walker1984satellite}
J.~G. Walker, ``Satellite constellations,'' \emph{Journal of the British Interplanetary Society}, vol.~37, p. 559, 1984.

\bibitem{AGI}
``Software for digital mission engineering and systems analysis,'' \url{https://www.ansys.com/products/missions/ansys-stk}, Jan. 2023.

\bibitem{9944663}
M.~Elmahallawy, T.~Elfouly, A.~Alouani, and A.~M. Massoud, ``A comprehensive review of lithium-ion batteries modeling, and state of health and remaining useful lifetime prediction,'' \emph{IEEE Access}, vol.~10, pp. 119\,040--119\,070, 2022.

\bibitem{helber2019eurosat}
P.~Helber, B.~Bischke, A.~Dengel, and D.~Borth, ``Eurosat: A novel dataset and deep learning benchmark for land use and land cover classification,'' \emph{IEEE Journal of Selected Topics in Applied Earth Observations and Remote Sensing}, vol.~12, no.~7, pp. 2217--2226, 2019.

\bibitem{xie2019asynchronous}
C.~Xie, S.~Koyejo, and I.~Gupta, ``Asynchronous federated optimization,'' \emph{arXiv preprint arXiv:1903.03934}, 2019.

\end{thebibliography}
\end{document}